\documentclass[12pt]{article}
\usepackage{a4wide,epsfig,psfrag,amsmath,amssymb,cite,scalefnt}
\usepackage{color}

\newcommand{\gsim}{\;\rlap{\lower 3.5 pt \hbox{$\mathchar \sim$}} \raise 1pt
 \hbox {$>$}\;}
\newcommand{\lsim}{\;\rlap{\lower 3.5 pt \hbox{$\mathchar \sim$}} \raise 1pt
 \hbox {$<$}\;}

\begin{document}

\title{\vskip-3cm{\baselineskip14pt
    \begin{flushleft}
      \normalsize SFB/CPP-11-36\\
      \normalsize TTP11-18
  \end{flushleft}}
  \vskip1.5cm
  Heavy quark and gluino potentials to two loops
}

\author{
  Tim Collet and
  Matthias Steinhauser
  \\[1em]
  {\small\it Institut f{\"u}r Theoretische Teilchenphysik}\\
  {\small\it Karlsruhe Institute of Technology (KIT)}\\
  {\small\it 76128 Karlsruhe, Germany}
}

\date{}

\maketitle

\thispagestyle{empty}

\begin{abstract}
  The static potentials for systems of a heavy quark and anti-quark,
  two gluinos and a quark and a gluino are computed for all possible colour
  configurations of a $SU(N_c)$ gauge group.
\medskip


\end{abstract}

\thispagestyle{empty}




The potential energy between two heavy quarks is one of the fundamental
quantities of the strong interaction and has been in the focus of the
theoretical investigations already in the early days of
QCD~\cite{Appelquist:1974zd}. The potential arises in a natural way when
considering the non-relativistic limit of a heavy quark and anti-quark
system as an ingredient of the resulting Schr\"odinger-like equations
(see Ref.~\cite{Brambilla:2004wf} for a review).
Thus, the potential constitutes a crucial input whenever the
production of heavy particles is considered at threshold or bound state
properties are calculated. Examples of Standard Model processes are the
production of top quark pairs in electron positron collisions for a
center-of-mass energy in the vicinity of twice the mass or the
invariant-mass distribution of $t\bar{t}$ pairs at hadron
colliders.  Furthermore, one should also mention the evaluation
of the energy levels and corrections to the wave function for heavy quark
bounds states like the $\Upsilon$ or $\Psi$ systems.

As far as processes beyond the Standard Model are concerned there have been
recent publications where bound states of two gluinos, the massive super
partners of the gluons, have been examined. Again, the corresponding
potential, which has been used to two-loop order, plays a crucial
role~\cite{Hagiwara:2009hq,Kauth:2009ud}. Similarly, in
Ref.~\cite{Kauth:2011-2} the threshold production of a gluino-squark pair is
considered. The required potential can be obtained from the quark-gluino
potential which is discussed below.

In this Letter we systematically compute the potentials of all colour
configurations of a quark-anti-quark, gluino-gluino and quark-gluino bound
state. To be precise, we consider the heavy-particle systems given in
Tab.~\ref{tab::qq} and
compute the potentials for the corresponding colour decomposition.

\begin{table}[htb]
  \begin{center}
    \renewcommand{\arraystretch}{1.5}
    \begin{tabular}{c|c|c}
      bound state & colour representation & irreducible representations\\
      \hline
      $q\bar{q}$ & $3\otimes\bar{3}$ & $1 \oplus 8$ \\
      $\tilde{g}\tilde{g}$ & $8\otimes 8$ &
      $1 \oplus 8_S\oplus 8_A\oplus 10\oplus\overline{10}\oplus 27\oplus R_7$ \\
      $q\tilde{g}$ & $3\otimes 8$ & $3 \oplus \overline{6} \oplus 15$ \\
    \end{tabular}
    \caption{\label{tab::qq}Heavy-particle systems and their colour
      decomposition into irreducible representations. The subscripts ``S'' and
      ``A'' distinguish the symmetric and anti-symmetric octet representations.}
  \end{center}
\end{table}

Note that in our framework both the heavy quark $q$ and gluino $\tilde{g}$ are
treated as external static colour sources added to the (massless) dynamical degrees of
freedom of QCD. Thus except for colour there is no difference in the treatment
of the gluino and the quark. As a consequence the potential of an anti-quark
and a gluino is identical to the $q\tilde{g}$ potential.

One comment
concerning the colour decomposition of the $\tilde{g}\tilde{g}$ potential is
in order: As it is common practice we consider only the
combination of $10\oplus \overline{10}$.
Furthermore, the colour structure $R_7$ is
only non-vanishing for $N_c\not=3$ and thus it is not relevant for
QCD~\cite{Bartels:1993ih,Nikolaev:2005zj,Dokshitzer:2005ig,Kovner:2005qj,Cvitanovic:2008zz}. 

We define the various potentials introduced above as follows
\begin{eqnarray}
  V_{ij}^{[c]}(\mu^2=\vec{q}\,^2) &=& -
  C^{[c]}\frac{4\pi\alpha_s(\vec{q}\,^2)}{\vec{q}\,^2}
  \left[1+\frac{\alpha_s(\vec{q}\,^2)}{4\pi}a_1
    +\left(\frac{\alpha_s(\vec{q}\,^2)}{4\pi}\right)^2\left(a_2 + \delta
      a_{2,ij}^{[c]}\right)
  \right]
  \,,
  \label{eq::pot}
\end{eqnarray}
where $ij\in\{q\bar{q},\tilde{g}\tilde{g},q\tilde{g}\}$ and $c$ defines the
colour state as given in Tab.~\ref{tab::qq}.
The renormalization scale is set to $\mu^2=\vec{q}\,^2$ to suppress the
trivial renormalization group terms on the r.h.s. of Eq.~(\ref{eq::pot}).
For the potential $V_{\tilde{g}\tilde{g}}^{[10]}$ we have to modify
Eq.~(\ref{eq::pot}) slightly since there is no tree and one-loop
contribution. Thus we write
\begin{eqnarray}
  V_{\tilde{g}\tilde{g}}^{[10]}(\mu^2=\vec{q}\,^2) &=& -
  \frac{4\pi\alpha_s(\vec{q}\,^2)}{\vec{q}\,^2}
  \left(\frac{\alpha_s(\vec{q}\,^2)}{4\pi}\right)^2
  \delta a_{2,\tilde{g}\tilde{g}}^{[10]}
  \,.
  \label{eq::pot10}
\end{eqnarray}

In Eq.~(\ref{eq::pot}) the coefficients
$a_1$ and $a_2$ are the one- and two-loop corrections
which are already present in the singlet contribution of the $q\bar{q}$
potential. They have been computed in
Refs.~\cite{Fischler:1977yf,Billoire:1979ih,Peter:1996ig,Peter:1997me,Schroder:1998vy}
and can be found in Ref.~\cite{Smirnov:2008pn} including higher order terms in
$(d-4)$ (where $d$ is the space-time dimension). The three-loop coefficient
$a_3$ has been computed in
Refs.~\cite{Smirnov:2008pn,Smirnov:2009fh,Anzai:2009tm}.
In less than four dimensions the static potential has recently been studied in
Ref.~\cite{Pineda:2010mb}, see also~\cite{Schroder:1999sg}, and the $N=4$
supersymmetric Yang-Mill theory has been considered in Ref.~\cite{Pineda:2007kz}.

At tree-level and at one-loop order the only difference among the various
potentials is due to the overall colour factor. At two-loop order we have
introduced the quantity
$\delta a_{2,ij}^{[c]}$ which parametrizes the difference to the singlet
result. It is currently only known for
$V_{q\bar{q}}^{[8]}$~\cite{Kniehl:2004rk}.
Furthermore, also for $V_{\tilde{g}\tilde{g}}^{[1]}$ the two-loop corrections
have been computed~\cite{Kauth:2009ud,Anzai:2010td} with the result $\delta
a_{2,\tilde{g}\tilde{g}}^{[c]}=0$. (In Ref.~\cite{Anzai:2010td} also the
three-loop term of $V_{\tilde{g}\tilde{g}}^{[1]}$ has been evaluated.) In this
Letter we present the two-loop results for all remaining potentials listed in
Tab.~\ref{tab::qq}.

\begin{figure}[t]
  \centering
  \epsfxsize=\textwidth
  \leavevmode
  \epsffile[65 315 560 530]{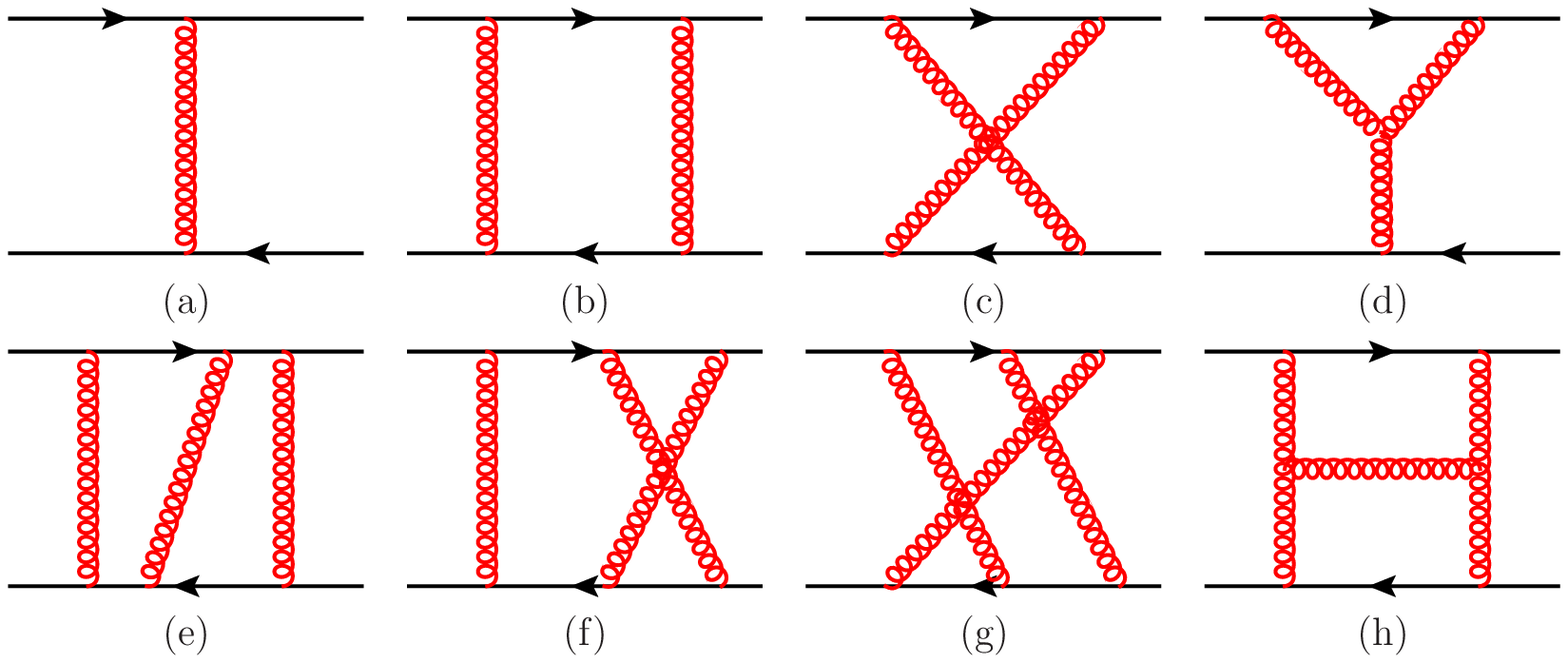}
  \caption[]{\label{fig::diags}Sample diagrams contributing to
    $V_{ij}^{[c]}$ at tree-level (a), one-loop (b)--(d) and
    two-loop order (e)--(h). 
    The straight lines correspond to quarks or gluinos,
    respectively, and the curly lines represent gluons.
    }
\end{figure}

For the calculation we have employed standard techniques which include the
automatic generation of the diagrams (see Fig.~\ref{fig::diags}), the
classification into different 
families of integrals, the application of
projectors~\cite{Bartels:1993ih,Nikolaev:2005zj,Dokshitzer:2005ig,Kovner:2005qj,Cvitanovic:2008zz}
and the reduction to master integrals using the Laporta
algorithm~\cite{Laporta:1996mq,Laporta:2001dd,FIRE}.  The latter have been
taken over from Ref.~\cite{Smirnov:2003kc}. The colour factors have been
computed with the help of the program {\tt color}~\cite{vanRitbergen:1998pn}.
We have performed the calculation for general gauge parameter and have checked
that it drops out in the final result.

The standard techniques for the evaluation of the loop integrals appearing at
one and two loops (see, e.g., Refs.~\cite{Schroder:1999sg,Smirnov:2003kc}) can
only be applied in a straightforward way to the singlet case since
there, apart from light-fermion contributions, only the maximally non-Abelian
parts contribute. In particular, no diagrams involving pinches occur, i.e. the
integrals do not contain propagators of the form $1/(k_0+i0)\times1/(k_0-i0)$
where $k$ is a loop momentum.  
Sample diagrams are shown in Figs.~\ref{fig::diags}(c) and (d) at
one-loop and Figs.~\ref{fig::diags}(g) and (h) at two-loop order.
However, for the non-trivial colour configurations also contributions
involving pinches (see, e.g., Figs.~\ref{fig::diags}(b), (e) and (f))
have to be taken into account. The results of the corresponding
diagrams are contained in the quantity $\delta a_{2,ij}^{[c]}$.
We evaluate the integrals by either exploiting the exponentiation of
the colour singlet potential or by carefully evaluating the potential
in coordinate space starting from the Wilson loop definition.
Both methods are described in detail
in Refs.~\cite{Schroder:1999sg,Kniehl:2004rk}. In this Letter we have checked
that they lead to the same result.

\begin{table}[t]
  \begin{center}
    \renewcommand{\arraystretch}{1.5}
    \begin{tabular}{c|c|c|c|c}
      $ij$ & $c$ & $C^{[c]}$ & $\delta a_{2,ij}^{[c]}$
      & $\delta a_{2,ij}^{[c]} (N_c=3)$ \\
      \hline
      $q\bar{q}$
      & 1 & $\frac{(N_c^2-1)}{2N_c}$ & 0 & 0 \\
      & 8 & $-\frac{1}{2N_c}$        & $N_c^2 \left(\pi^4-12\pi^2\right)$ 
      & $-189.2$ \\
      \hline
      $\tilde{g}\tilde{g}$
      & 1 & $N_c$ & 0 & 0\\
      & $8_S$ & $ \frac{N_c}{2} $    & 0 & 0\\ 
      & $8_A$ & $ \frac{N_c}{2} $    & $ - 6  \left(\pi^4-12\pi^2\right)$ 
      & $126.2$ \\
      & $10$ &  ---   & $-\frac{3 N_c}{2}
      \left(\pi^4-12\pi^2\right)$ 
      & $94.62$ \\
      & 27 & $ -1 $    & $\frac{1}{2} \left( N_c + 2 \right) \left(
      N_c + 1 \right) \left(\pi^4-12\pi^2\right)$ 
      & $-210.3$ \\
      & $R_7$ & $ 1 $    & $\frac{1}{2} \left( N_c - 2 \right) \left(
      N_c - 1 \right) \left(\pi^4-12\pi^2\right)$ & --- \\
      \hline
      $q\tilde{g}$
      & 3 & $\frac{N_c}{2}$ & $- \left(\pi^4-12\pi^2\right)$ & $21.03$ \\
      & $\overline{6}$ & $ \frac{1}{2}  $   & $\frac{1}{2} N_c \left(
      N_c - 3 \right) \left(\pi^4-12\pi^2\right)$ & 0 \\
      & 15 & $ - \frac{1}{2}  $   & $\frac{1}{2} N_c \left( N_c + 3
      \right) \left(\pi^4-12\pi^2\right)$ & $-189.2$\\ 
    \end{tabular}
    \caption{\label{tab::res}Results for the $C^{[c]}$ and $\delta
      a_{2,ij}^{[c]}$ for the various colour configurations. In the right
    column we set $N_c=3$ and evaluate $\delta a_{2,ij}^{[c]}$ numerically.
    Note that $c=10$ refers to the combination $10\oplus\overline{10}$.
    Furthermore, $R_7$ has dimension zero for $N_c=3$.}
  \end{center}
\end{table}

In Tab.~\ref{tab::res} we present our results for 
$C^{[c]}$ and $\delta a_{2,ij}^{[c]}$ for $SU(N_c)$.
Note that $\delta a_2$ is zero for the singlet contributions but also
for gluino bound states in the symmetric octet
configuration. One furthermore obtains a vanishing result for the
representation $\overline{6}$
($q\tilde{g}$) when specifying to QCD, i.e., setting $N_c=3$.
It is remarkable that all non-vanishing contributions are
proportional to the same combination $(\pi^4-12\pi^2)$ with a prefactor
depending on the colour state
although the individual diagrams contributing to 
$\delta a_{2,ij}^{[c]}$ do not show this proportionality and
furthermore also contain terms without $\pi^2$ or $\pi^4$.

As far as the numerical importance of $\delta a_2$ is concerned one
can compare the results in the last column of Tab.~\ref{tab::res} with
$a_2$ for the bottom and top system given by
\begin{eqnarray}
  a_2(n_l=4) &\approx& 211.1\,,\nonumber\\
  a_2(n_l=5) &\approx& 155.8\,.
\end{eqnarray}
(The corresponding numbers for $a_1$ are $5.889$ and $4.778$,
respectively.)
In some cases one observes a significant reduction of the two-loop
coefficient (see, e.g., the $27$ configuration for $n_l=4$ where 
$a_2+\delta a_{2,\tilde{g}\tilde{g}}^{[10]}\approx0.8$) 
whereas in other cases the
large value of $a_2$ is even further increased.

To conclude, in this Letter the quark-anti-quark, gluino-gluino and
quark-gluino potentials have been computed for all possible colour
configurations up to two loops. In all cases it is possible to
identify the two-loop coefficient $a_2$ originating from the
quark-anti-quark singlet potential. The additional contributions
are given by a colour factor times $(\pi^4-12\pi^2)$.


\section*{Acknowledgements}

We would like to thank Matthias Kauth, 
Johann K\"uhn and Alexander Penin for many useful discussions and
communications.
This work was supported by the Deutsche Forschungsgemeinschaft through
the SFB/TR-9 ``Computational Particle Physics''.



\end{document}